\documentclass[reprint,notitlepage,showpacs,preprintnumbers,nofootinbib,amsmath,amssymb,aps,prd,floatfix]{revtex4-1}
\pdfoutput=1
\usepackage{graphicx}
\usepackage{bm}
\usepackage{subfigure}
\usepackage{url}
\usepackage[hyperindex]{hyperref}
\usepackage{color}
\usepackage[ddmmyy,24hr]{datetime}
\usepackage{bigdelim}
\usepackage{booktabs}
\usepackage{dcolumn}
\usepackage{multirow}
\usepackage{subfigure}
\usepackage{cancel}
\usepackage{stackrel}
\usepackage{paralist}
\usepackage{xspace}
\usepackage{slashed}
\newcommand{\nua}[1]{\ensuremath{\rlap{\kern-2.5pt\ensuremath{\overset{\scriptscriptstyle(-)}{\phantom{\nu}}}}{\ensuremath{{\nu}_{#1}}}}}

\usepackage{enumitem}
\begin{document}

\title{Statistical Significance of Reactor Antineutrino Active-Sterile Oscillations}

\author{C. Giunti}
\email{carlo.giunti@to.infn.it}
\affiliation{Istituto Nazionale di Fisica Nucleare (INFN), Sezione di Torino, Via P. Giuria 1, I--10125 Torino, Italy}

\date{15 May 2020}

\begin{abstract}
We performed Monte Carlo calculations of the
statistical distribution of the $\chi^2$ test statistic
used in the analysis of the data of the
NEOS,
DANSS,
Bugey-3, and
PROSPECT
short-baseline reactor experiments.
We show that the statistical significance of the
NEOS and DANSS indications in favor of active-sterile neutrino oscillations
is smaller than that obtained with the usual method based on the $\chi^2$ distribution.
In the combined analysis of the data of the four experiments
we find that the statistical significance of
active-sterile neutrino oscillations is reduced from
$2.4\sigma$
to
$1.8\sigma$.
\end{abstract}


\maketitle

Active-sterile neutrino oscillations due to a nonstandard massive neutrino at the eV scale
is one of the current hot topics of research that could reveal the existence
of new physics beyond the Standard Model.
This line of research is historically motivated by three indications
of short-baseline neutrino oscillations that cannot be explained in the standard
three-neutrino mixing framework
(see the recent reviews in Refs.~\cite{Giunti:2019aiy,Diaz:2019fwt,Boser:2019rta}):
1) the LSND observation of short-baseline $\bar\nu_{\mu}\to\bar\nu_{e}$ transitions \cite{Aguilar:2001ty}
(the LSND anomaly);
2) the measurement of short-baseline $\nu_e$ disappearance
\cite{Laveder:2007zz,Giunti:2006bj}
in the gallium source experiments
GALLEX
\cite{Kaether:2010ag}
and
SAGE
\cite{Abdurashitov:2005tb}
(the gallium neutrino anomaly);
3) the short-baseline $\bar\nu_e$ disappearance of reactor $\bar\nu_e$
\cite{Mention:2011rk}
with respect to the theoretical prediction of the reactor $\bar\nu_e$ fluxes
\cite{Mueller:2011nm,Huber:2011wv}
(the reactor antineutrino anomaly).
In particular, the discovery in 2011 of the reactor antineutrino anomaly
raised a great interest and triggered the realization of several new
reactor experiments searching for short-baseline $\bar\nu_e$ disappearance
in a model-independent way,
that does not depend on the theoretical prediction of the reactor $\bar\nu_e$ fluxes.
This goal can be achieved by confronting the antineutrino detection rates
or energy spectra measured at different distances from the reactor
and fitting the ratio with the effective short-baseline (SBL) survival probability
\begin{equation}
P_{ee}^{\text{SBL}}
=
1 - \sin^2\!2\vartheta_{ee} \, \sin^2\!\left( \dfrac{\Delta{m}^2_{41} L}{4 E} \right)
,
\label{Pee}
\end{equation}
where
$\Delta{m}^2_{41}=m_{4}^2-m_{1}^2 \gtrsim 1 \, \text{eV}^2$
is the difference between the squared mass of the nonstandard neutrino $\nu_{4}$
and the squared masses
$ m_{1}, m_{2}, m_{3} \ll m_{4} $
of the standard neutrinos in the 3+1 active-sterile mixing scheme,
that is the simplest one that can explain short-baseline neutrino oscillations.
The amplitude of the oscillations is given by
$\sin^2\!2\vartheta_{ee} = 4 |U_{e4}|^2 ( 1 - |U_{e4}|^2 )$,
where $\vartheta_{ee}$ is the effective mixing angle
for short-baseline $\nu_{e}$ and $\bar\nu_e$ disappearance
and
$U$ is the $4\times4$ unitary neutrino mixing matrix.
Different experiments are characterized by different ranges of
the neutrino energy $E$ and measure the neutrino
interaction rate with detectors located at different distances $L$
from the neutrino source.

Of particular importance are the results of the
NEOS~\cite{Ko:2016owz}
and
DANSS~\cite{Alekseev:2018efk,Danilov:2019aef}
experiments,
that have a remarkable agreement for oscillations generated by
$ \Delta{m}^2_{41} \approx 1.3 \, \text{eV}^2 $
and
$ \sin^2\!2\vartheta_{ee} \approx 0.02 - 0.05 $
\cite{Gariazzo:2018mwd,Dentler:2018sju,Berryman:2019hme,Giunti:2019fcj}.
The statistical significance of the combined NEOS and DANSS
indication of short-baseline oscillations
is $3.7\sigma$~\cite{Gariazzo:2018mwd}
using the published 2018 DANSS results~\cite{Alekseev:2018efk}
and $2.6\sigma$~\cite{Giunti:2019fcj}
using the preliminary 2019 DANSS results~\cite{Danilov:2019aef}.
This statistical significance is clearly not
enough to claim a discovery,
especially taking into account the decrease
passing from the 2018 to the 2019 DANSS results.
However, the indication is intriguing
and it is worth careful examination.

A source of misinterpretation of neutrino oscillation experiments
is the assumption that the $\chi^2$ test statistic
(or twice the negative logarithm of the likelihood)
considered in the analysis of the data
follows a $\chi^2$ distribution,
according to Wilks' theorem~\cite{Wilks:1938dza}.
This is often not correct and can lead to significant statistical
inaccuracies
\cite{Feldman:1997qc,Lyons:2014kta,Agostini:2019jup,Algeri:2019arh}.
In particular,
as explained in Ref.~\cite{Agostini:2019jup},
the analysis of binned data in terms on neutrino oscillations
suffers from serious undercoverage for values of the mixing
that are below the sensitivity of the experiment.
This is due to the fact that even in the absence of real oscillations,
binned data can often be fitted better by oscillations
that reproduce the statistical fluctuations of the bins.
In order to overcome this problem,
some experimental collaborations analyze their data
performing a Monte Carlo estimation of the distribution of the
$\chi^2$ test statistic.
More precisely,
in order to evaluate properly the allowed regions in the
oscillation parameter space it is necessary to evaluate by Monte Carlo
the distribution of the difference $\Delta\chi^2$ between the $\chi^2$
in each point of the parameter space and the global minimum of the $\chi^2$.
This is an expensive computational task that is typically performed on a grid
in the two-dimensional oscillation parameter space, that involves tens of thousands of points.
Instead,
assuming the $\chi^2$ distribution for $\Delta\chi^2$,
the contours of the allowed regions
in the oscillation parameter space are simply given by
the fixed values of $\Delta\chi^2$
in Table~39.2 of Ref.~\cite{Tanabashi:2018oca} for the number of degrees of freedom
equal to number of oscillation parameters.

So far, the combined analyses of the data of different experiments,
sometimes called ``global fits'',
have been performed assuming the $\chi^2$ distribution
(except for the analysis of solar neutrino data in Ref.~\cite{Garzelli:2000yf}),
for two main reasons:
\begin{enumerate}

\renewcommand{\labelenumi}{(\theenumi)}
\renewcommand{\theenumi}{\Alph{enumi}}

\item
A Monte Carlo estimation of the distribution of the
$\Delta\chi^2$ would require a detailed knowledge of each experiment.

\item
The computational task for evaluating
the distribution of the
$\Delta\chi^2$
on a fine grid in the parameter space is prohibitive.

\end{enumerate}

In particular,
the combined analyses of the data of NEOS, DANSS, and other reactor experiments
in terms of short-baseline oscillations
\cite{Gariazzo:2018mwd,Dentler:2018sju,Berryman:2019hme,Giunti:2019fcj}
have been performed assuming the $\chi^2$ distribution.
This can lead to a serious overestimation of the statistical significance
of the indication of short-baseline oscillations.

In this paper we improve the combined analysis of
NEOS, DANSS, and other reactor experiments
presented in Refs.~\cite{Gariazzo:2018mwd,Giunti:2019fcj}
by performing the required Monte Carlo estimation of the distribution of the
$\Delta\chi^2$.
We overcame the two difficulties A) and B) mentioned above by:
\begin{enumerate}

\renewcommand{\labelenumi}{(\theenumi)}
\renewcommand{\theenumi}{\alph{enumi}}

\item
Extracting from the published experimental papers all the information
that allows one to analyze the experimental data taking into account the
uncertainties that are shown in the figures
or provided in supplementary files.
This approach neglects the correlations
of the systematic uncertainties of binned data
when it is not publicly available.
Therefore, the analysis is not ``exact'',
but it is an acceptable approximation for the estimation of the distribution of the
$\Delta\chi^2$.

\item
Performing the Monte Carlo simulations
on a reduced grid of points in the two-dimensional
($\sin^2\!2\vartheta_{ee},\Delta{m}^2_{41}$)
plane with a few thousands of points
and interpolating the distribution of the $\Delta\chi^2$.
This a good approximation,
because the variations of the distribution of the $\Delta\chi^2$ are smooth.

\end{enumerate}

\begin{figure}[!t]
\centering
\includegraphics*[width=\linewidth]{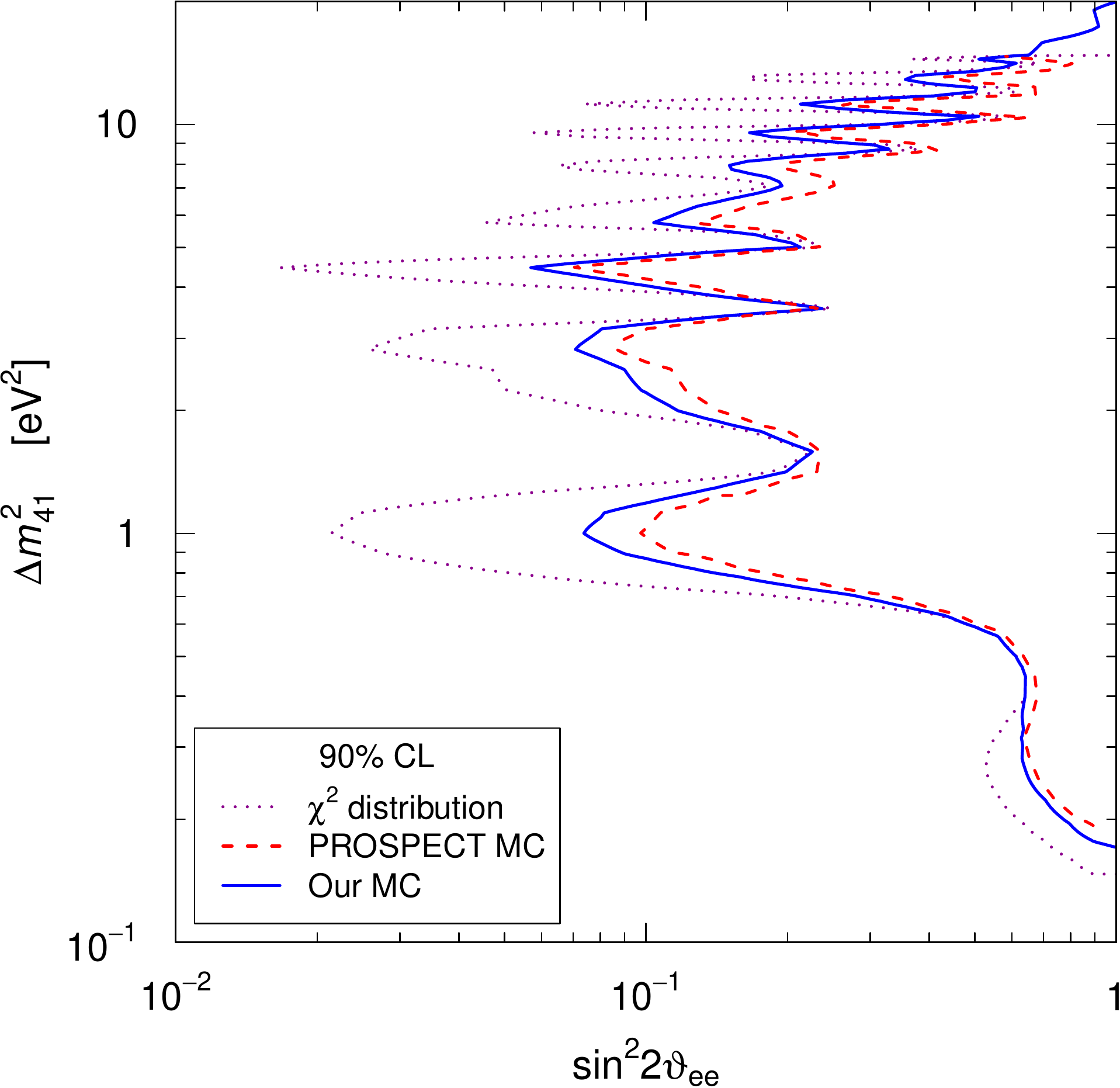}
\caption{ \label{fig:prospect90}
90\% CL exclusion curves in the
($\sin^2\!2\vartheta_{ee},\Delta{m}^2_{41}$)
plane obtained with the $\chi^2$ distribution for $\Delta\chi^2$ (magenta dotted),
with the PROSPECT $\Delta\chi^2$ map (red dashed),
and with our Monte Carlo calculation (blue solid).
}
\end{figure}

Regarding the item a), let us further emphasize that
the mentioned approximation is adopted only
for the estimation of the distribution of the
$\Delta\chi^2$
with Monte Carlo simulated data.
On the other hand,
the evaluation of the value of the $\chi^2$ for the actual data
must take into account the correlations
of the systematic uncertainties of binned data,
if they exist and are known.
For the NEOS experiment we use the $\chi^2$ table
kindly provided by the NEOS collaboration,
that takes into account all uncertainties.
For the DANSS experiment we use a $\chi^2$ without bin correlations,
as done by the DANSS collaboration~\cite{Alekseev:2018efk,Danilov:2019aef}.

Besides the data of the NEOS and DANSS experiments,
we took into account also the spectral-ratio data of the
Bugey-3~\cite{Declais:1995su}
and
PROSPECT~\cite{Ashenfelter:2018iov}
experiments\footnote{
We cannot include in the analysis the results of the STEREO~\cite{Almazan:2018wln,AlmazanMolina:2019qul}
and
Neutrino-4~\cite{Serebrov:2018vdw,Serebrov:2020rhy}
experiments, because there is not enough available information.
},
that exclude large values of $\sin^2\!2\vartheta_{ee}$
in a model-independent way,
as shown in Ref.~\cite{Giunti:2019fcj}.
For Bugey-3 we consider the 40/15 m ratio of positron spectra in Figure~15 of Ref.~\cite{Declais:1995su}
and the $\chi^2$ in Eq.~(8) of the same paper.
This $\chi^2$ does not contain bin correlations,
for which there is no information in Ref.~\cite{Declais:1995su}.
For PROSPECT, we used the published $\Delta\chi^2$ map~\cite{Ashenfelter:2018iov}
that takes into account all the uncertainties.

The analysis of the PROSPECT data is a
useful test of our method,
because the PROSPECT collaboration published the
map in the
($\sin^2\!2\vartheta_{ee},\Delta{m}^2_{41}$)
plane
of the $\Delta\chi^2$ corresponding to
90\% and 95\% CL,
that they evaluated with a Monte Carlo that takes into account
all the experimental details,
including the correlations
of the systematic uncertainties of the binned data.
Figure~\ref{fig:prospect90} shows the comparison of the
90\% CL exclusion curves in the
($\sin^2\!2\vartheta_{ee},\Delta{m}^2_{41}$)
plane obtained with the $\chi^2$ distribution for $\Delta\chi^2$ (magenta dotted),
with the PROSPECT 90\% CL $\Delta\chi^2$ map (red dashed),
and with our Monte Carlo calculation (blue solid).
One can see that our method yields an exclusion curve that is close to the
PROSPECT one and can be considered as an acceptable approximation,
especially taking into account the large improvement with respect to the
exclusion curve obtained assuming the $\chi^2$ distribution.

\begin{figure*}[!t]
\centering
\setlength{\tabcolsep}{0pt}
\begin{tabular}{cc}
\subfigure[]{\label{fig:neos}
\begin{tabular}{c}
\includegraphics*[width=0.49\linewidth]{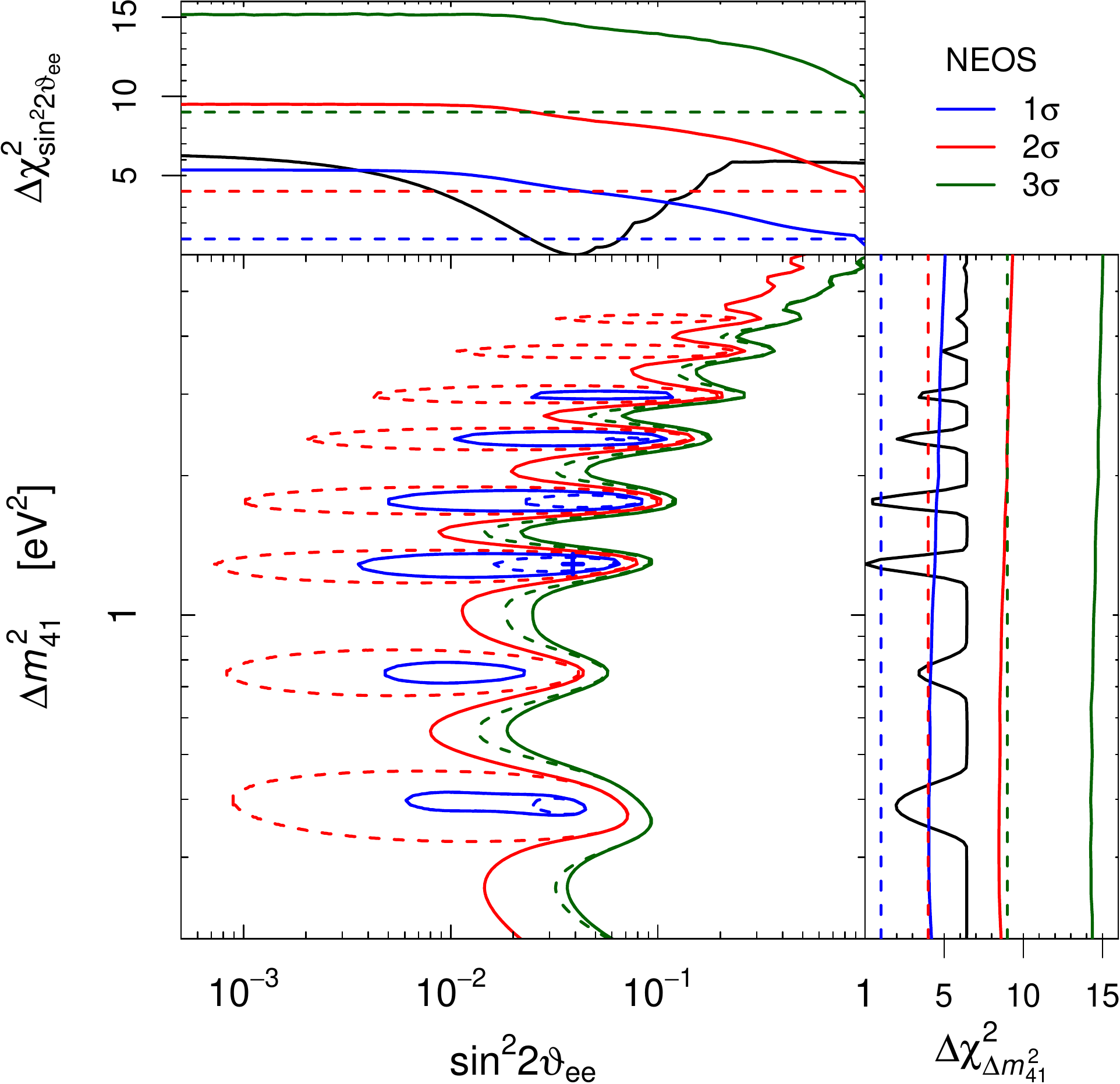}
\\
\end{tabular}
}
&
\subfigure[]{\label{fig:danss}
\begin{tabular}{c}
\includegraphics*[width=0.49\linewidth]{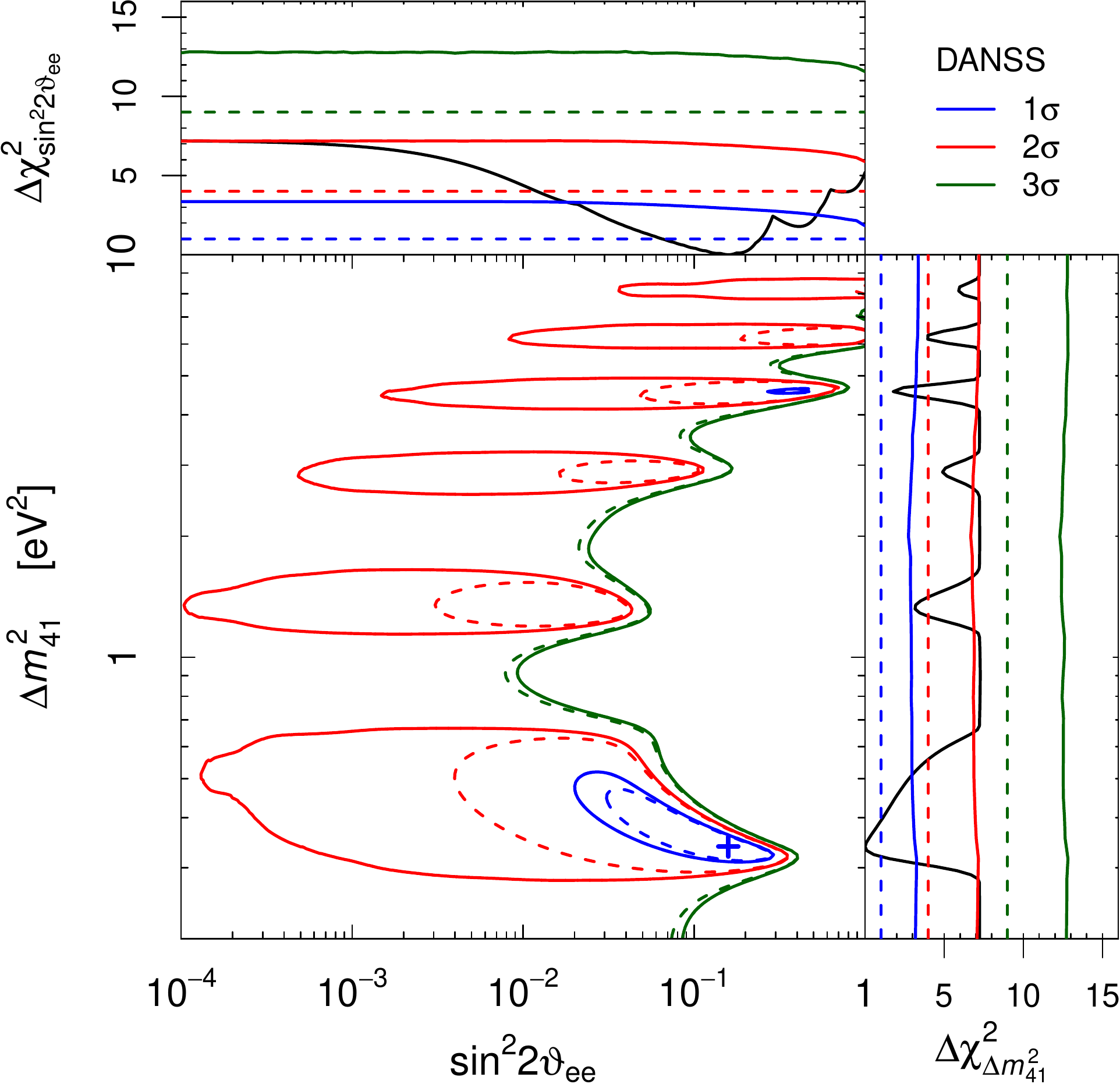}
\\
\end{tabular}
}
\\
\subfigure[]{\label{fig:bugey}
\begin{tabular}{c}
\includegraphics*[width=0.49\linewidth]{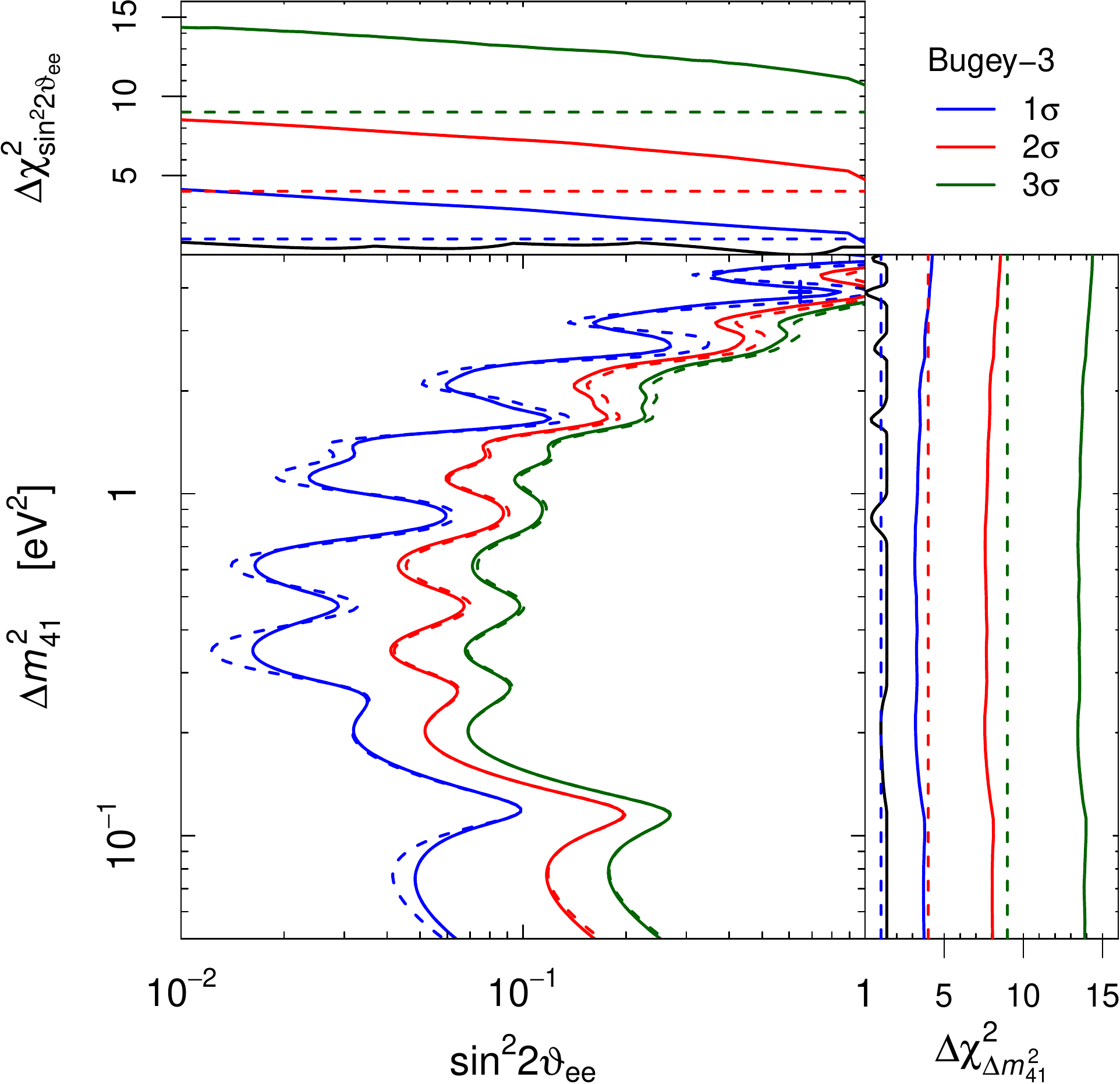}
\\
\end{tabular}
}
&
\subfigure[]{\label{fig:prospect}
\begin{tabular}{c}
\includegraphics*[width=0.49\linewidth]{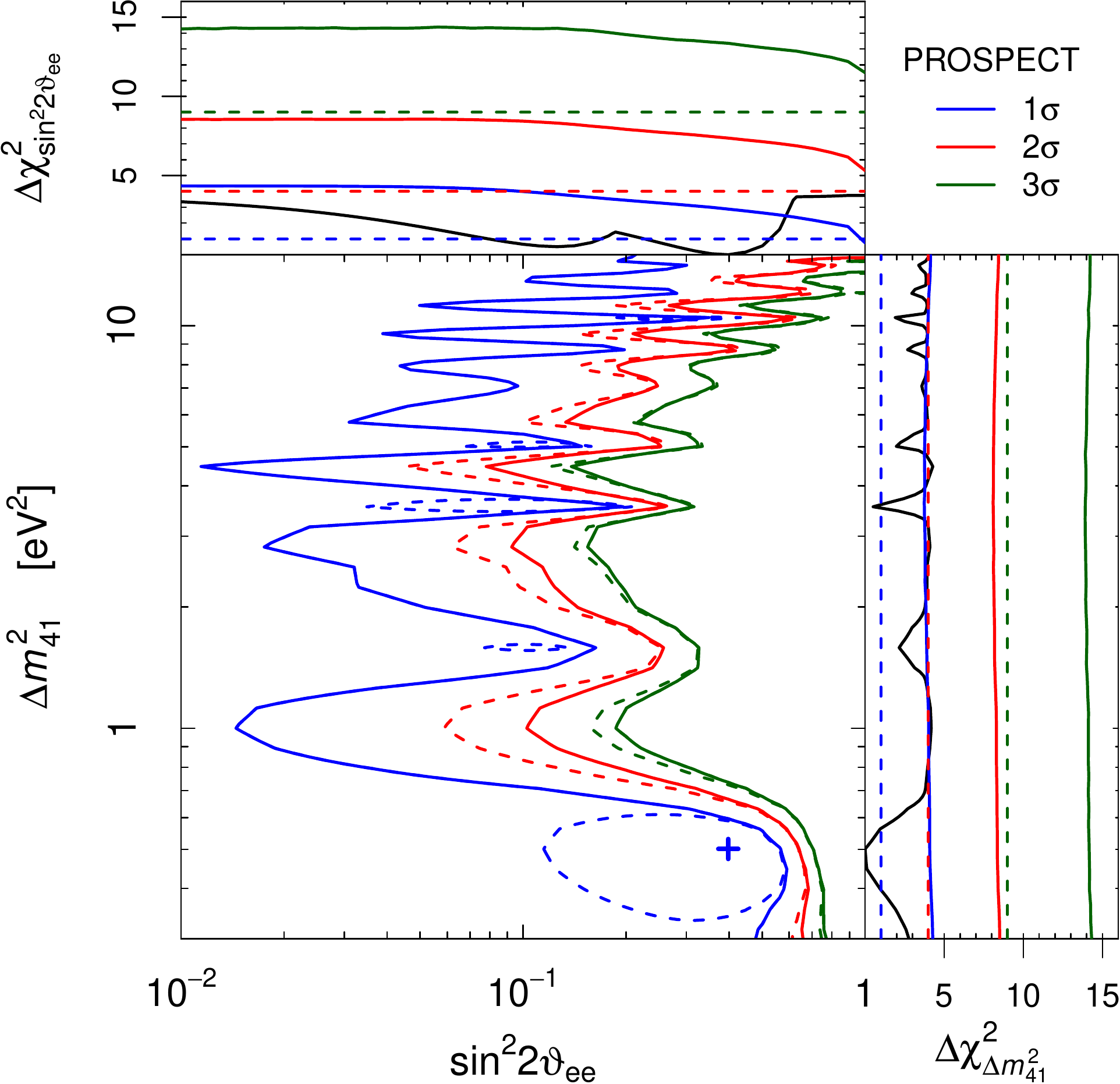}
\\
\end{tabular}
}
\end{tabular}
\caption{ \label{fig:four}
Contours of the
$1\sigma$ (blue),
$2\sigma$ (red), and
$3\sigma$ (green)
allowed regions in the
($\sin^2\!2\vartheta_{ee},\Delta{m}^2_{41}$)
plane
for the four reactor spectral-ratio experiments
\subref{fig:neos}
NEOS~\cite{Ko:2016owz},
\subref{fig:danss}
DANSS~\cite{Danilov:2019aef},
\subref{fig:bugey}
Bugey-3~\cite{Declais:1995su}, and
\subref{fig:prospect}
PROSPECT~\cite{Ashenfelter:2018iov}.
The solid lines represent the contours obtained with our Monte Carlo evaluation of
the distribution of $\Delta\chi^2$,
and the dashed lines depict the contours obtained assuming the $\chi^2$ distribution.
Also shown are the marginal $\Delta\chi^2$'s (black) for
$\sin^2\!2\vartheta_{ee}$ and $\Delta{m}^2_{41}$,
together with the $\Delta\chi^2$ values corresponding to
$1\sigma$ (blue),
$2\sigma$ (red), and
$3\sigma$ (green) obtained with the $\chi^2$ distribution (dashed)
and our Monte Carlo (solid).
The best-fit points are indicated by blue crosses.
}
\end{figure*}

Having established the approximate reliability of our method,
we present in Figure~\ref{fig:four} the contours of the
$1\sigma$,
$2\sigma$, and
$3\sigma$
allowed regions in the
($\sin^2\!2\vartheta_{ee},\Delta{m}^2_{41}$)
plane
for the four reactor spectral-ratio experiments that we take into account:
NEOS~\cite{Ko:2016owz},
DANSS~\cite{Danilov:2019aef},
Bugey-3~\cite{Declais:1995su}, and
PROSPECT~\cite{Ashenfelter:2018iov}.
The figures show also the marginal $\Delta\chi^2$'s for
$\sin^2\!2\vartheta_{ee}$ and $\Delta{m}^2_{41}$
and the values of the marginal $\Delta\chi^2$'s corresponding to
$1\sigma$,
$2\sigma$, and
$3\sigma$
obtained with the $\chi^2$ distribution (dashed)
and our Monte Carlo (solid).
Note that the marginal $\Delta\chi^2$'s corresponding to
different confidence levels obtained with the $\chi^2$ distribution
have fixed values given by Table~39.2 of Ref.~\cite{Tanabashi:2018oca} for one degree of freedom,
whereas the Monte Carlo ones depend on the value of the unmarginalized parameter.

From Figure~\ref{fig:four}
one can see that in general the difference between the
Monte Carlo contours and those obtained with the $\chi^2$ distribution
is larger for small values of $\sin^2\!2\vartheta_{ee}$,
where the $\chi^2$ distribution leads to undercoverage,
as discussed in Ref.~\cite{Agostini:2019jup}.
This is due to the larger probability of fitting the statistical fluctuations of the data
with fake oscillations when the signal is small or absent.
Correspondingly,
the Monte Carlo values of the marginal $\Delta\chi^2_{\sin^2\!2\vartheta_{ee}}$'s
associated with different confidence levels
increase in a significant way for decreasing $\sin^2\!2\vartheta_{ee}$.
On the other hand,
the Monte Carlo values of the marginal $\Delta\chi^2_{\Delta{m}^2_{41}}$'s
corresponding to different confidence levels are approximately flat,
with slightly smaller values in the range of $\Delta{m}^2_{41}$
where the experiment is mostly sensitive.
However,
they are significantly larger than those obtained with the $\chi^2$ distribution,
because for each value of $\Delta{m}^2_{41}$
the contribution to the marginalization of small values of $\sin^2\!2\vartheta_{ee}$
lead to undercoverage in the case of the $\chi^2$ distribution.
In the range of $\Delta{m}^2_{41}$
where the experiment is mostly sensitive,
the undercoverage is slightly smaller, because it occurs only for very small values of
$\sin^2\!2\vartheta_{ee}$.

\begin{table}[!t]
\begin{center}
\begin{tabular}{lccccc}
\\
&
&
\multicolumn{2}{c}{$\chi^2$ distrib.}
&
\multicolumn{2}{c}{MC distrib.}
\\
&
$\Delta\chi^2_{\text{NO}}$
&
$p$-value
&
$n\sigma$
&
$p$-value
&
$n\sigma$
\\
\hline
NEOS
&
$6.4$
&
$0.041$
&
$2.0$
&
$0.2$
&
$1.3$
\\
DANSS
&
$7.2$
&
$0.027$
&
$2.2$
&
$0.044$
&
$2.0$
\\
Bugey-3
&
$1.4$
&
$0.5$
&
$0.7$
&
$0.9$
&
$0.1$
\\
PROSPECT
&
$4.0$
&
$0.14$
&
$1.5$
&
$0.37$
&
$0.9$
\\
Combined
&
$8.5$
&
$0.015$
&
$2.4$
&
$0.079$
&
$1.8$
\\
\hline
\end{tabular}
\caption{ \label{tab:pvl}
Value of $\Delta\chi^2_{\text{NO}}$
and the associated $p$-value
and number of $\sigma$'s
obtained with the $\chi^2$ distribution
and with the Monte Carlo distribution.
The lines correspond to the four reactor spectral-ratio experiments
NEOS~\cite{Ko:2016owz},
DANSS~\cite{Danilov:2019aef},
Bugey-3~\cite{Declais:1995su},
PROSPECT~\cite{Ashenfelter:2018iov},
and to the combined fit.
}
\end{center}
\end{table}

\begin{figure}[!t]
\centering
\includegraphics*[width=\linewidth]{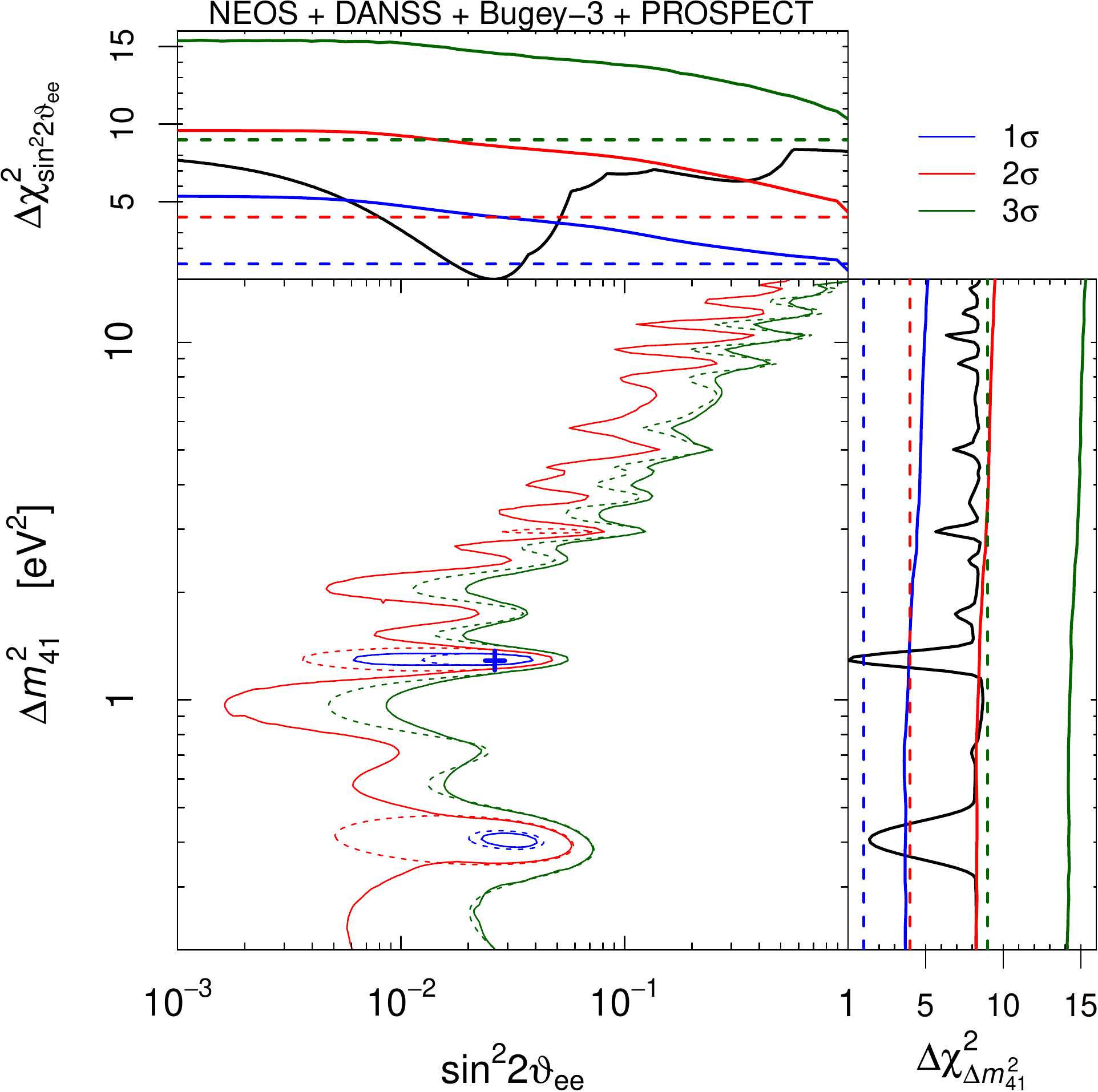}
\caption{ \label{fig:combined}
Contours of the
$1\sigma$ (blue),
$2\sigma$ (red), and
$3\sigma$ (green)
allowed regions in the
($\sin^2\!2\vartheta_{ee},\Delta{m}^2_{41}$)
plane
obtained with the combined analysis of the data of the
four reactor spectral-ratio experiments
NEOS~\cite{Ko:2016owz},
DANSS~\cite{Danilov:2019aef},
Bugey-3~\cite{Declais:1995su}, and
PROSPECT~\cite{Ashenfelter:2018iov}.
The solid lines represent the contours obtained with our Monte Carlo evaluation of
the distribution of $\Delta\chi^2$,
and the dashed lines depict the contours obtained assuming the $\chi^2$ distribution.
Also shown are the marginal $\Delta\chi^2$'s (black) for
$\sin^2\!2\vartheta_{ee}$ and $\Delta{m}^2_{41}$,
together with the $\Delta\chi^2$ values corresponding to
$1\sigma$ (blue),
$2\sigma$ (red), and
$3\sigma$ (green) obtained with the $\chi^2$ distribution (dashed)
and our Monte Carlo (solid).
The blue cross indicates the best-fit point.
}
\end{figure}

The Monte Carlo evaluation of the distribution of the
$\Delta\chi^2$
allows us also to estimate reliably the statistical significance
of the indication in favor of active-sterile neutrino oscillations in each experiment.
This is done by considering the difference
$ \Delta\chi^2_{\text{NO}} = \chi^2_{\text{NO}} - \chi^2_{\text{min}} $,
where $\chi^2_{\text{NO}}$ is the value of the $\chi^2$
obtained without neutrino oscillations.
If the conditions for Wilks' theorem~\cite{Wilks:1938dza}
hold, $\Delta\chi^2_{\text{NO}}$ has a $\chi^2$
distribution with two degrees of freedom
corresponding to the two fitted parameters
$\sin^2\!2\vartheta_{ee}$
and
$\Delta{m}^2_{41}$.
Table~\ref{tab:pvl}
shows the values of $\Delta\chi^2_{\text{NO}}$ that we obtained in the analyses of
the data of the four reactor spectral-ratio experiments
NEOS~\cite{Ko:2016owz},
DANSS~\cite{Danilov:2019aef},
Bugey-3~\cite{Declais:1995su}, and
PROSPECT~\cite{Ashenfelter:2018iov}.
Also shown are the $p$-value of the no-oscillation hypothesis
and the corresponding number of $\sigma$'s for a two-sided Gaussian deviation
obtained with the $\chi^2$ distribution
and
the Monte Carlo distribution.
The statistical significance of the indication in favor of active-sterile neutrino oscillations
can be quantified by the smallness of the $p$-value or by the number of $\sigma$'s.
One can see that for all the considered experiments this
statistical significance is reduced considering the Monte Carlo distribution of
$\Delta\chi^2_{\text{NO}}$
with respect to the $\chi^2$ distribution.

The NEOS indication of neutrino oscillations is strongly reduced from
$2.0\sigma$
to
$1.3\sigma$
using the Monte Carlo distribution instead of the $\chi^2$ distribution.
Note that our Monte Carlo
$20\%$
$p$-value is in very good approximate agreement with the 22\%
$p$-value estimated by the NEOS collaboration~\cite{Ko:2016owz}.
This is another confirmation of the validity of our method.

Using the Monte Carlo distribution instead of the $\chi^2$ distribution,
the significance of the DANSS indication of neutrino oscillations
is only slightly reduced
from
$2.2\sigma$
to
$2.0\sigma$.

From Table~\ref{tab:pvl} one can also see that
the Bugey-3 experiment has clearly no indication of neutrino oscillations,
independently of the assumed distribution of $\Delta\chi^2_{\text{NO}}$.
Indeed, although the best-fit point in Figure~\ref{fig:bugey}
corresponds to large oscillations,
all the confidence level contours are exclusion curves
and the values of the marginal $\Delta\chi^2$'s are very small.

The PROSPECT $\chi^2$ has a minimum for large oscillations,
as shown by Figure~\ref{fig:prospect},
but the value of the $\Delta\chi^2$ is small and
in agreement with the absence of oscillations.
The Monte Carlo calculation of the distribution of the $\Delta\chi^2$
gives a statistical significance for oscillations
that is only
$0.9\sigma$,
and significantly smaller than the
$1.5\sigma$
obtained with the $\chi^2$ distribution.
Our Monte Carlo $p$-value of
$37\%$
is not as large as the 58\% obtained by the PROSPECT collaboration~\cite{Ashenfelter:2018iov},
but it is much closer to it than the
$14\%$
obtained with the $\chi^2$ distribution.

We finally performed the combined fit of the data of all the four
reactor spectral-ratio experiments
NEOS~\cite{Ko:2016owz},
DANSS~\cite{Danilov:2019aef},
Bugey-3~\cite{Declais:1995su}, and
PROSPECT~\cite{Ashenfelter:2018iov}.
Figure~\ref{fig:combined} shows the contours of the
$1\sigma$,
$2\sigma$, and
$3\sigma$
allowed regions in the
($\sin^2\!2\vartheta_{ee},\Delta{m}^2_{41}$)
plane and the marginal $\Delta\chi^2$'s for
$\sin^2\!2\vartheta_{ee}$ and $\Delta{m}^2_{41}$.
As in Figure~\ref{fig:four},
the dashed lines correspond to the $\chi^2$ distribution,
whereas the solid lines have been obtained with our Monte Carlo.
The combined fit yields the best-fit point at
$
\sin^2\!2\vartheta_{ee}
=
0.026
$
and
$
\Delta{m}^2_{41}
=
1.3
\, \text{eV}^2
$
with
$
\chi^2_{\text{min}}
=
183.6
$
and
$197$
degrees of freedom.
Therefore,
the fit is very good,
either considering a $\chi^2$ distribution for $\chi^2_{\text{min}}$
($75\%$ goodness of fit)
or with a Monte Carlo estimation
($89\%$ goodness of fit).
The best fit corresponds to the allowed regions of
NEOS and DANSS at
$
\Delta{m}^2_{41}
\approx
1.3
\, \text{eV}^2
$,
that can be seen in Figures~\ref{fig:neos} and~\ref{fig:danss}.

Figure~\ref{fig:combined} shows that the
Monte Carlo allowed regions are much larger than those
obtained with the $\chi^2$ distribution
and extend to smaller values of the mixing.
This means that the preference for neutrino oscillations is smaller.
Indeed, from Table~\ref{tab:pvl} one can see that
the significance of the indication of neutrino oscillations
is reduced from
$2.4\sigma$
to
$1.8\sigma$.
Correspondingly,
in Figure~\ref{fig:combined} there are several closed contours that delimit
regions allowed at 90\% CL,
that corresponds to about $1.64\sigma$.

In conclusion,
our Monte Carlo calculations show that the
significance of the indication of neutrino oscillations
given by the combined analysis of the
reactor spectral-ratio measurements
is not as large as thought
before~\cite{Gariazzo:2018mwd,Dentler:2018sju,Berryman:2019hme,Giunti:2019fcj}
using a $\chi^2$ distribution for $\Delta\chi^2$.
Nevertheless,
there is still an marginal
$1.8\sigma$
indication in favor of active-sterile neutrino oscillations
that will hopefully be resolved by new data of the ongoing
DANSS~\cite{Alekseev:2018efk,Danilov:2019aef},
PROSPECT~\cite{Ashenfelter:2018iov},
STEREO~\cite{Almazan:2018wln,AlmazanMolina:2019qul},
Neutrino-4~\cite{Serebrov:2018vdw,Serebrov:2020rhy}, and
SoLid~\cite{Abreu:2020bzt}
reactor spectral-ratio experiments.

The interpretation of the results of the experiments searching
for active-sterile neutrino oscillations is currently controversial.
In this situation it is important that the combined analyses of the data of different experiments
are performed by calculating the appropriate distribution of the test statistic,
as we have done here,
in order to reach reliable information on the statistical significance of active-sterile neutrino oscillations.
The quality of such analysis depends on how well the test statistic
takes into account the experimental setup
and the systematic uncertainties, including their correlation.
To this aim it is necessary that each experimental collaboration
make available all the necessary information.

\begin{acknowledgments}
This work was partially supported by the research grant
``The Dark Universe: A Synergic Multimessenger Approach'' number 2017X7X85K
under the program ``PRIN 2017'' funded by the Ministero dell'Istruzione,
Universit\`a e della Ricerca (MIUR).
\end{acknowledgments}

%

\end{document}